 \documentclass[twocolumn,showpacs,preprintnumbers,amsmath,amssymb]{revtex4}

\usepackage{graphicx}
\usepackage{dcolumn}
\usepackage{bm}

\begin{document}

\preprint{}

\title{Polymer-Induced Bundling of F-actin and the Depletion Force}

\author{ M. Hosek}
\affiliation{
Physics Department \\
Indiana University \\
Bloomington, Indiana 47405 \\
}
\author{ J. X. Tang }
\affiliation{
Physics Department \\
Brown University \\
Providence, Rhode Island 02192\\
}


\date{\today}

\begin{abstract}

  The inert polymer polyethylene glycol (PEG)
induces a "bundling" phenomenon in F-actin solutions when its concentration exceeds
a critical onset value $C_o$.
Over a limited range of PEG molecular weight and ionic strength, $C_o$ can be expressed as 
a function of these two variables.
The process is reversible, but hysteresis is also observed in the dissolution of the bundles, 
with ionic strength having a large influence.
Additional actin filaments are able to join previously formed
bundles.
PEG polymers are not incorporated
into the actin bundles.

  Estimates of the Asakura-Oosawa depletion force, Coulomb
repulsion, and van der Waals potential are combined in order to explain 
 the bundling effect and hysteresis.
Conjectures are presented concerning the apparent limit in bundle size. 

\end{abstract}


\maketitle



\section{Introduction}

Cells and tissues are crowded with macromolecules such as proteins, DNA, 
and various polymers. While the ligand-substrate specific 
binding model forms the central dogma of biochemistry, many protein-protein 
interactions are facilitated in large part by physical drives such as the 
excluded volume effect, electrostatic interaction, and depletion force. Chemically 
inert polymers such as polyethylene glycol (PEG) and dextran are often added into 
solutions of biomolecules in order to mimic the crowded biochemical environment 
and understand various biomolecular functions ranging from protein filament assembly~\cite{Ito89},
ion channel opening and closing~\cite{parsegian-pnas}, to transcription of DNA~\cite{rau-pnas}. 
This study focuses on the 
physical effects of PEG on the lateral aggregation of a filamentous protein assembly, 
with the goal of a 
first principle explanation of its aggregation property. Similar phenomena 
occur in many cellular and physiological settings, which by large are dictated by 
the common physical 
mechanisms, although often in more complicated and less defined conditions.

The protein selected for this study is actin. 
Actin is a ubiquitous cytoskeletal protein of molecular weight (MW) 42000 Dalton. 
In solutions of low ionic
strength it exists as a globular monomer, G-actin. As 
$[\rm{K^+}]$ or $[\rm{Na^+}]$ is increased beyond 50~mM, the monomers polymerize into
helical filaments (37~nm pitch) of side-by-side monomers incorporated into two 
strands~\cite{prot_prof, holmes_90},
known as filamentous actin, or F-actin. F-actin is polydisperse in length, and consists of
370 monomers per $\mu$m unit length. The diameter of F-actin is 8~nm. 
At pH~8, the linear charge density is
about 4~e/nm and surface charge density 0.15 e/nm$^2$.
This value is derived from the amino acid sequence of $\alpha$-skeletal 
muscle actin, where each monomer has a net charge of $-12e$~\cite{prot_prof, JXT96}.
In a dilute solution F-actin
is a freely undulating filament with a bending modulus $k_c$ such that the
persistence length ${L_p = k_c/kT} = 17~\rm{\mu m} $~\cite{Gittes_93,Venier_95}. 
This length is larger than an averaged filament length of F-actin. In the study of lateral 
aggregation of F-actin as induced by 
flexible polymers such as PEG, the effect of
flexibility of F-actin is negligible, and the filaments are treated as charged colloidal rods.

 The physical insight into the interaction of free polymers with colloidal
particles was first achieved by Asakura and Oosawa (AO) a half century ago~\cite{AandO54}. 
The essential prediction from the AO treatment is that an attractive force is generated between 
two colloidal particles in the presence of non-interacting polymers. The effect, 
known widely as the depletion force, is essentially of entropic origin, and has been 
calculated for various sizes and geometries 
of the colloids~\cite{UIUC_2000,AALouis2002b,AALouis2002a,tuinier_2002,Schafer99}.

\begin{figure}
\includegraphics{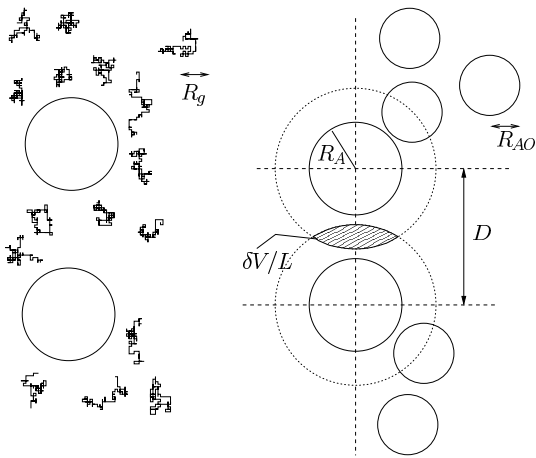}   
 \caption{
 \label{fig:rod-rod}
 Cross-section of two aligned cylinders with a depletion,or exclusion, layer.
$R_A$ is the F-actin radius, $R_g$ is the PEG radius of gyration, and $D$ is
the axis-to-axis separation of the actin cylinders. 
On the left is a cartoon of the two cylinders surrounded by
PEG, represented as self-avoiding random walks on a cubic lattice, N=180,
corresponding to PEG~8000~\cite{UIUC_2000}.
In this case $C$ is well below $C^*$.
In the AO scheme on the right, the colloidal rods are surrounded by
interpenetrating hard spheres of effective radius $R_{AO}$.
The attractive interaction is $P \delta V$, where $ \delta V$ is the
product of the hatched overlap area and an arbitrary length $L$. 
}
 \end{figure}

To apply the AO model, flexible polymers are typically 
treated as freely interpenetrating hard spheres of radius $R_{AO}$, which are excluded from the colloid surface 
by a thin layer of thickness $R_{AO}$.  
It was shown~\cite{AandO54} that this shell creates a positive free energy difference,
$\Delta F = P V = P R_{AO} A_c $, where $P$ is the osmotic pressure due to
the polymer, and $A_c$ the surface area of the colloid. 
If two colloidal particles share part of this volume, the volume 
accessible to the polymers is increased. Consequently, the total entropy of the system is 
increased and thus the free energy of the system is lowered. In terms of
the Helmholtz free energy,
\begin{equation}
\Delta F(V,T) = {{{\partial F} \over {\partial V}} \delta V} = P  \delta V. \label{AOeqn}
\end{equation}
For an ideal gas, $ {{\partial {S}} \over {\partial {V}}}  = {{P}\over{T}} $,
thus it is clear that the AO interaction is entropically driven.
However, osmotic pressure and volume of exclusion are more accessible
to measurement and  calculation.
The geometry of the colloids and their depletion layer is portrayed in 
Fig~\ref{fig:rod-rod}.
The volume of the exclusion layer is  a function of the polymer radius of gyration $R_g$, 
the radius of the colloidal rods $R_A$, and 
the axis-to-axis separation between the rods $D$.
It has been shown~\cite{AALouis2002b} for an ideal random chain polymer that 
$R_{AO} =2 R_g/\sqrt{\pi}$, where $R_g$ is the radius of gyration.

 \hfill
 \newline
 \hfill
 \newline

 Polyethylene glycol (PEG) has long been used as an agent for precipitation 
and crystallization 
of proteins.  PEG is a polymer that finds water a "good" solvent, in that
it assumes the configuration of an almost ideal self-avoiding random walk~\cite{selser91},
with
radius of gyration $R_g \sim MW^{3/5}$~\cite{deGennes79book}.
As the
polymer concentration $C$ exceeds a value $C^*$, the
molecules begin to overlap, and direct scaling arguments~\cite{deGennes79book}  show that
the correlation length $\xi$ begins to decrease as  
$({C}/{C^*})^{-3/4}$.
The thickness of the polymer depletion layer at the colloid surface~\cite{AALouis2002a}  
scales with  $\xi$, with 
the polymer monomer number density continuously passing
from zero at the colloid surface to its bulk value. The 
nature of this profile is the subject of
continuing investigation.
Nevertheless, in a continuum approach Eqn.~\ref{AOeqn} becomes
a volume integral, $ \Delta F = \int P(\bf{r}) \rm{d}V $.

  Two analytical treatments have been proposed recently to calculate the correlation 
length $\xi$ and osmotic pressure $P$ in the semidilute regime. 
The renormalization group theory~\cite{Schafer99}
yields
expressions~\cite{tuinier_2001,oono82}
for $\xi$ and $P$
as functions of polymer concentration and molecular weight. 
A second treatment has been 
developed by Schweizer and co-workers~\cite{UIUC_2000,chandler_88},  which employs the 
polymer reference 
interaction site model (PRISM). The numerical results from these two
treatments agree with each other 
in a range of concentrations spanning the dilute and semidilute regimes.
The predictions also agree well with the available experimental data.  
Therefore, results from both treatments are applied later in this work in 
calculating the depletion force.

 Interacting charged colloids in saline solution are typically modeled as bodies with Coulomb
repulsion and attractive short-range dispersion forces, a combination known as
the DLVO theory~\cite{overbeek_book,israelachvili}.
The electrostatic interaction in this situation is modeled 
with the  Poisson-Boltzmann equation,
which describes the relation between charge density $\rho$ and electrical potential $\phi$,
under the assumption that the ionic charges deviate from their bulk concentration $n_0$ 
according to the Boltzmann law. For monovalent ions:

\begin{equation}
\nabla^2 \phi = - \rho / \epsilon \nonumber
\end{equation}

\begin{equation}
\rho = e(n_+ - n_-) = {n_0 e }[{e^{-{\phi e} / {kT} } - e^{{\phi e} / {kT} } }] \nonumber
\end{equation}

or

\begin{equation}
\nabla^2 \phi ={2 n_0 e  \sinh ( {{\phi e} / {kT} })} / {\epsilon}. \nonumber
\end{equation}
With boundary conditions of moderate surface charge, the
solutions of this equation are associated with 
a characteristic exponential decay length known as the  Debye screening 
length  $\kappa ^{-1}$, which for monovalent ions is
proportional to $1/\sqrt{n_0}$. 

  Dispersion forces are the result of mutually induced dipole attraction 
between two bodies, and as such follow a $1/r^6$ law for interaction potential~\cite{overbeek_book},
and are of course shorter in range than the electrostatic interaction.
At high enough salt concentration, the 
electrostatic repulsion between colloidal particles of like charge can become weaker than 
the attractive dispersion force. As a result,  the 
colloids precipitate. 

The rest of the paper is organized as follows. Section II describes materials and methods
employed for this work. Section III presents
quantitative data and microscopic observations 
about the formation of F-actin bundles that result when the PEG concentration 
exceeds a critical onset value $C_o$.  
Much of the analytical effort is made in section IV, by employing what are basically
estimates of the electrostatic repulsion, van der Waals
attractions (the DLVO theory), and the depletion effect. Qualitative explanations are made
concerning onset of bundle formation, hysteresis, 
as well as the bundle size limit. Limitations 
of the present treatment and the experiments performed are also discussed.

\section{Materials and Methods}

\subsection {Materials}
 G-actin was prepared from acetone powder of rabbit skeletal muscle after
Spudich and Watt~\cite{SpWatt71}
and stored as 0.2~mL aliquots,  7.9~g/L, at -80 $^o$C until use.
G-actin was polymerized by first diluting to 3.2 g/L with 
"G-buffer", and then adding 3~M KCl to bring [KCl] to 150~mM.
Polymerization to F-actin was immediately evident by a rapid increase in 
viscosity. Polymerization was always allowed to proceed for 
several hours.
"G-buffer" is 0.2~mM $\rm{CaCl_2}$, 0.5~mM ATP, 0.5~mM $\rm{NaN_3}$, 0.5~mM DTT, 2.0~mM tris-HCl at pH 8.0.
To create stock solutions where $\rm{[KCl] > 0.3~M}$, the salt was added directly to G-buffer.
PEG of molecular weights 4000 to  20000 were purchased from Sigma 
and PEG~35000 from Alfa-Aesar. 
Radiolabeled PEG was obtained from American Radiolabeled Chemicals (ARC-1565, $\rm{^3H}$ -labeled PEG~35000)
and  $\rm{ ^{14}C }$ labeled PEG 4000 from Amersham Biosciences, Inc. (CFA508).

\subsection {Light Scattering Measurement of $C_o$}
  Ninety degree light scattering was measured by adopting a Perkin-Elmer LS-5B luminescence 
spectrophotometer~\cite{JXT96} using a 1~mL culture tube (4.5~mm ID) to hold the 
sample. The tube was aligned with the probe beam for the least refraction  by
the glass tube and maximum illumination of the sample. Monochromatic light of 550~nm wavelength
was used 
for illumination and detection, with spectral slit width selected from 3 to 15~nm to keep the 
detected light in the linear range of the photomultiplier. Once a measurement was 
commenced by adding the initial PEG for a certain KCl concentration, the spectrophotometer
settings were kept unchanged for consistency.

  Concentrated PEG solution was typically added in volumes
of  $\rm{5~\mu L}$ for a typical increment of 
0.2\% in the w/w percentage of PEG. 
Thorough mixing was achieved using a long, 0.2~mL pipet tip with a cut end ($\rm{\sim}$~1~mm~dia),
and shear was minimized by keeping the flow rate below 0.2~mL/1~sec. 
During the first couple of mixing cycles, the contrast in the
index of refraction in the mixture was seen to quickly diminish. This observation 
suggests that the mixture became
homogeneous following several cycles of pipeting.
F-actin concentration
was 0.3 g/L with a sample volume of 0.4~mL. 
$C_o$ was characterized by a sudden rise in turbidity, developing in less than 
60 seconds, and a corresponding increase in light scattering.

\subsection {The  amount of PEG in the bundle structure }
Approximately 2000 Bq (2000 counts/s) of $\rm{^{3}H}$ labeled PEG~35000 was added to 0.7~mL F-buffer, 
600~mM KCl, in a
glass test tube of 1~cm diameter. The sample was then mixed with 0.2~mL of 3.2~g/L F-actin.
To induce bundling, 10~$\mu$L increments of 13\% w/w unlabeled PEG~35000 in F-buffer 
containing 0.6~M~KCl
were added with thorough yet gentle mixing. When the light scattering
(turbidity)  characteristic of bundling was observed, one more  10~$\mu$L aliquot was added,
and the mixture was allowed to sit for 10 min. This resulting w/w concentration of PEG
solution for the onset of bundling is denoted $C_o$. 
Then the entire sample was 
transfered to a 1.5~mL plastic Eppendorf vial and centrifuged at 1200~g for 5~min. in 
a swinging bucket rotor. 
The supernate $S0$ was extracted 
and the  pellet  was immediately resuspended in $\rm{100~\mu L}$ of 7\% w/w unlabeled 
PEG F-buffer containing 600~mM KCl.
At this point 1.2~mL more of the same unlabeled PEG buffer was added and any remaining clumps 
resuspended by pipet.
To minimize shear in resuspension, the flow rate was limited to $\rm{100~\mu L /sec}$ in pipeting,
and the pipet tip had the nozzle cut for a  0.5~mm opening.   
  
  After allowing the resuspension to re-equilibrate for 30 min, the protein was 
pelleted again at 8000~g for  5 min. in a swinging bucket rotor,
and the supernate $S1$ extracted.  
To minimize the pellet mass, the vial was again centrifuged at 13000~g for 5~min. in a 
fixed angle rotor, and the last remaining supernate removed with
capillary action.
The total weight of the vial was then obtained to $\rm{50~\mu g}$ accuracy, and the
pellet  weight $\rm{W_{pellet}}$ derived from the vial tare.
The pellet was then suspended in $\rm{100~\mu L} $ of  2\% SDS and quantitatively washed into 5~mL of
scintillation fluid (IRL BioSafe II). $\rm{10~\mu L}$ samples of $S0$ and $S1$ 
were added to 5~mL of scintillation fluid, and the activity measured with a Beckman LS6500 
scintillation counter.

\subsection{ Hysteresis  in bundle formation }
 For a given KCl concentration (100~mM or 600~mM) and 0.3~g/L F-actin, 
PEG 35000 was added in
0.2~\% increments, with thorough but gentle mixing. When the qualitative 
change in turbidity associated with bundle formation was observed, the
mixture was allowed to equilibrate for 30 minutes. Then,
the sample was split into 0.3~mL aliquots, and each centrifuged
at 2000~g. 
An appropriate volume was removed from the top,
and replaced with F-buffer to reduce the PEG concentration.
The bundles were then gently resuspended with a cut pipet
tip.
After 12 hours at 4~$^o$C, the bundles were pelleted at 2000~g~$\times$~7 min.
and the supernate removed. Centrifugation was repeated to remove remaining
traces of F-buffer. The bicinchoninic acid method (Sigma B-9643)
was used to assay the remaining bundled protein in the pellet.
DTT from the F-buffer was found to interfere with this assay, but
was not significant in relation to the final amount of measured protein.

\subsection{ Numerical solution of cylindrical PB equation }
 A 'shooting' technique~\cite{Num_Rec}, with potential $\phi$ as the independent
variable~\cite{wooding_70},  was applied to numerically generate solutions to the 
PB equation with boundary conditions for an infinite cylinder. 
Exploiting the symmetry of the problem, we set $\phi_i(r_i) = i \it{\Delta} \phi$ for $ i = 0,1, ...$.
Debye-Huckel theory tells us that the counter charge density $\rho$ between $r_i$ and
$r_{i+1}$ is $2 n_{_{0}} e \sinh({\bar{\phi}_{_{i}} e/kT}) / \epsilon$,
where $\bar{\phi}_{_{i}} = ({\phi}_{_{i}} + {\phi}_{_{i+1}})/2$.
Charge conservation is expressed with the relationship 
\begin{equation}
  \lambda_{i+1}  =  \lambda_i +  {\it{\Delta}} r  2 \pi r_i \rho(\bar{\phi_i}) \nonumber
\end{equation}
where $\lambda_i $ represents the net linear charge density enclosed within the \it{i}\rm th shell.  
This, with the following approximation of Gauss's Law
\begin{equation}
{{\it{\Delta} \phi} \over {\it{\Delta} r}} ={ {\lambda_{i+1} } \over { 2 \pi r_i \epsilon } } \nonumber
\end{equation}
allows a quadratic equation to be solved for $\it{\Delta} r$:
\begin{equation}
{\it{\Delta}}{ \phi = {\it{\Delta}} r ( {\it{\Delta}} r 2 \pi r_i \rho(\bar{\phi_i}) + \lambda_i)} / { 2 \pi r_i \epsilon  }  \nonumber
\end{equation}
After an initial guess of $r_0 = B$, $\phi_0 = 0$ for the outer boundary condition, 
$r_i$ is calculated iteratively.
If, at $r_i \sim R_{\rm{actin}}$, $\lambda_i \sim -\lambda_{\rm{actin}}$,
the solution set $r_i(\phi_i)$ is accepted. If not, a new $B$ is chosen.
A value of 0.1~mV = $kT/250$ is used for $\it{\Delta} \phi$. Typically the 
charge balance condition is satisfied to within 2~\%.

\section{Results}
\subsection{Threshold and reversibility of bundle formation}

Four parameters of  phase separation were explored: PEG MW, PEG concentration $C$,
ionic strength, and actin concentration. 
The solution pH was held fixed at 8.0.
As polymer concentration $C$ is increased, a
sudden increase in light scattering marks the concentration $C_o$ where 
the onset of bundling is induced~\cite{JXT97,JXT96,Ito89}. At monovalent ionic strength near 100~mM
 and PEG MW 8000 
or below, the
turbidity of the sample is low as the structures are small.
These small structures were also measured by fluorescent imaging (data not shown).
For 35~kD~PEG, the light scattering is stronger, as the
structures are much larger in both length and diameter. 

\begin{figure}
\includegraphics{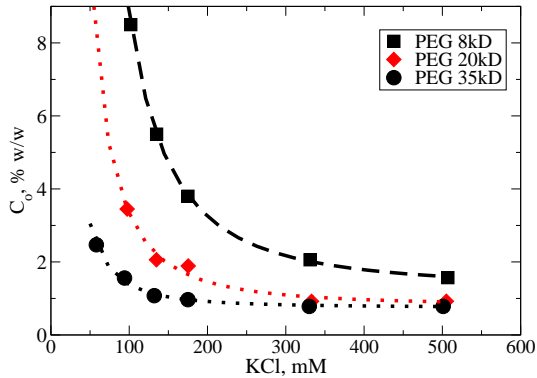}
 \caption{
 \label{fig:C_o}
 The effects of PEG MW and ionic strength on bundling  
critical concentration $C_o$, as detected by the onset of light 
scattering.
The lines through the points are semi-empirical fits using Eqn.~\ref{fit_eqn},
where $K_A$ = 1/(5~mM) for each data set,
but with an adjustable baseline $a_0$ for each MW.
F-actin concentration for this data set is 0.3g/L.
}
 \end{figure}

 The effects of PEG MW and the solution ionic strength on $C_o$ are shown in Fig~\ref{fig:C_o}.
The data clearly suggest an interplay of Debye screening length and polymer $R_g$.
These curves can be fit with the following semi-empirical function:
\begin{equation}
C_o = a_0 + a_1/\rm{(1+[KCl]/5~mM)}^2  \label{fit_eqn}
\end{equation}
where $C_o$ is percent w/w of PEG. 
$a_1$ is found to be a rather strong function of molecular weight. A powerlaw fit yields 
$a_1 = 70.0 {(MW/8000)}^{-2.0}$.
Separate measurements at 600~mM KCl show that $a_0$ is a weak function of molecular weight,
$a_0 = 1.0 {(MW/8000)}^{-0.5}$.
The term $1/\rm{(1+[KCl]/5~mM)}$ could be interpreted as 
the result of mass action ion association 
between the 
total bound charge of the F-actin $Q_0^-$ and the solution potassium ions
of concentration $[K^+]$:
\begin{equation}
\rm{Q^- + K^+ \stackrel{K_A}{\stackrel{\textstyle{\rightharpoonup}} {\leftharpoondown}} QK } \nonumber
\end{equation}
\begin{equation}
K_A = {{[QK]} \over {[K^+] [Q^-]} }  \label{K_A_eqn}
\end{equation}
Hence
\begin{equation}
 [Q^-] = [Q_0^-]/(1 + K_A [K^+])  \nonumber
\end{equation}
where $K_A =\rm{ 1/{(5~mM)}}$ is the association constant in this ligand binding model~\cite{dewey_90}.
One can speculate that an effective charge $[Q^-]$
dominates the electrostatic interaction
at lower ionic strength, where the repulsion between two F-actins, proportional to $[Q^-]^2$, 
is overcome by a depletion force
proportional to the osmotic pressure at the onset of bundling.
Similar empirical fits can be applied
to the mobility of fd virus as a function of monovalent and divalent ionic strength
[ Q. Wen and J.X. Tang, unpublished data].  

 \hfill
 \newline
 \hfill
 \newline

 Measurements show small effect of protein concentration
on $C_o$ over the range of 0.2~g/L to 0.8~g/L F-actin, although the 
kinetics for bundle formation, as indicated by light scattering,
were appreciably slower for the lowest F-actin concentration (data not shown).

 Also, the bundling process is reversible. After PEG concentration passes
$C_o$, the protein may be sedimented  by centrifugation and resuspended in F-buffer with
a resulting disappearance of turbidity.
PEG may then again be added to induce bundles.

 \hfill
 \newline
 \begin{figure}
\includegraphics{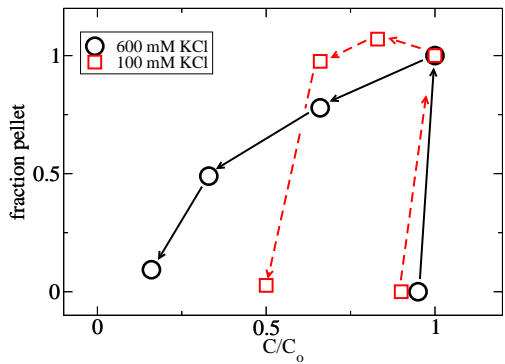}
 \caption{
 \label{fig:dissolve}
 Hysteresis in the dissolution of PEG 35000-induced bundles at 
[KCl] = 100~mM and 600~mM. The PEG concentration is normalized
to $C/{C_o}$, and the amount of recovered protein (bundles) is 
plotted as the fraction of the total protein. 
For $C \leq C_o$, there were virtually no bundles formed. As PEG is added,
almost all the F-actin went into bundles and was pelleted. After this, as
the PEG concentration was lowered in suspension, many of the
bundles did not dissolve until $C$ was lowered by over 50 \%.
}
 \end{figure}
 \hfill
 \newline

 Hysteresis is seen in the disappearance of the bundles
if the  polymer concentration is increased beyond $C_o$ and then, by dilution, decreased below $C_o$.
As shown in Fig ~\ref{fig:dissolve}, there is a pronounced effect of ionic strength, with bundles 
formed at a higher ionic strength more stable.

\subsection{Nucleation and growth}

\begin{figure*}  
\includegraphics{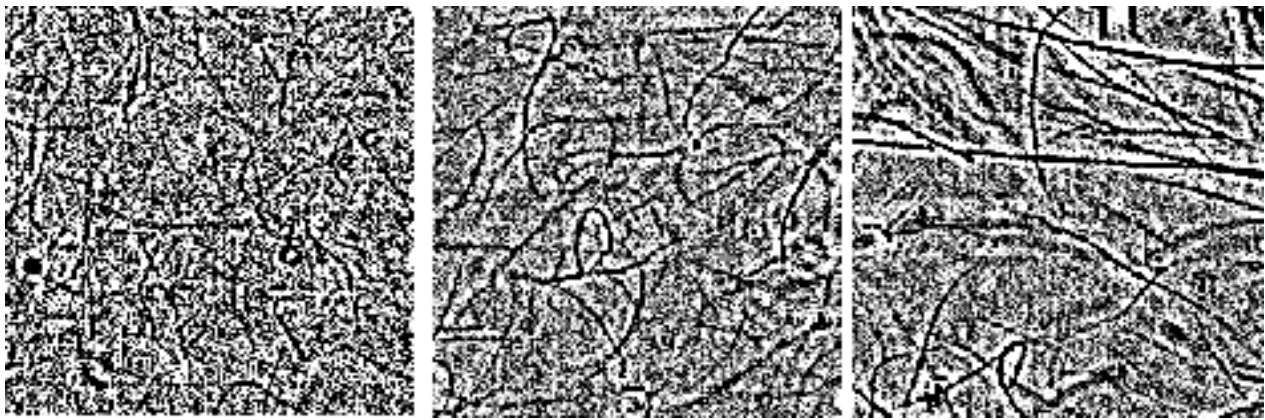}
 \caption{
 \label{fig:Nucl}
F-actin bundle formation and progression following  the gradual addition of
PEG~35000 to just past $C_o$.
The three representative pictures show bundles immediately after the
gradual addition of
PEG~35000 to just past $C_o$ (left panel), after 5~min.~(middle), and after
30~min.~(right). [KCl]~=~100~mM.
Images are  160 $\times$ 160~$\mu$m, taken
with a Nikon TE300 microscope, 10X phase contrast. The samples were
flowed into a channel of 70 $\mu$m thickness created by placing two pieces of double
sided tape between a microscope slide and a cover slip.
F-actin concentration is 0.2~g/L.
}
\end{figure*}

Immediately after the polymer concentration reaches $C_o$, 
actin filaments coalesce into thin strings (Fig~\ref{fig:Nucl}, left). 
After a few minutes, with mild intermittent mixing, the typical
size increases (Fig~\ref{fig:Nucl}, middle).
Then, with mild intermittent mixing over 30 min, 
these initial structures grow into polydisperse structures larger in diameter 
and length (Fig~\ref{fig:Nucl}, right). The bundles in the first image have roughly $10^5$ F-actin filaments 
each; those in the final image are at least an order of magnitude more massive. 
Therefore, some consideration must be given to the kinetics of 
bundle formation.  The series of images in Fig ~\ref{fig:Nucl} shows a 
typical time progression of the actin bundles induced by PEG~35000. 
It should be noted that there is no role of PEG concentration heterogeneity 
in the maturation, as the polymer concentration is entirely 
homogeneous after the mixing process completed within the first few seconds. 
It was found, however, through separate experiments,
that additional mixing at the later stages also affected 
the kinetics of the
bundle growth. The physical mechanism of this additional observation is clear in that 
as the bundle size increases, diffusion becomes progressively inefficient in facilitating
the self-assembly.

 When F-actin  is added to a solution where $C = 5 C_o$, 
large aggregates with sheet- or balloon-like structure are formed. 
This must largely be associated with the macroscopic,
slow mixing process of viscous media, easily seen as discontinuities
of index of refraction on the mm length scale.  
Furthermore, actin will coalesce without mixing of
the PEG solution. If a $\rm{10~\mu L}$ drop of F-actin is placed on
the surface of a denser 5\% PEG solution, the concentrated protein is found 
as one mass on the surface.

 Under all circumstances, there were no signs of slow spinodal decomposition or 
coexistence of phases.

 \hfill  
 \newline
 \hfill
 \newline

 \hfill
 \newline
 \begin{figure*}
\includegraphics{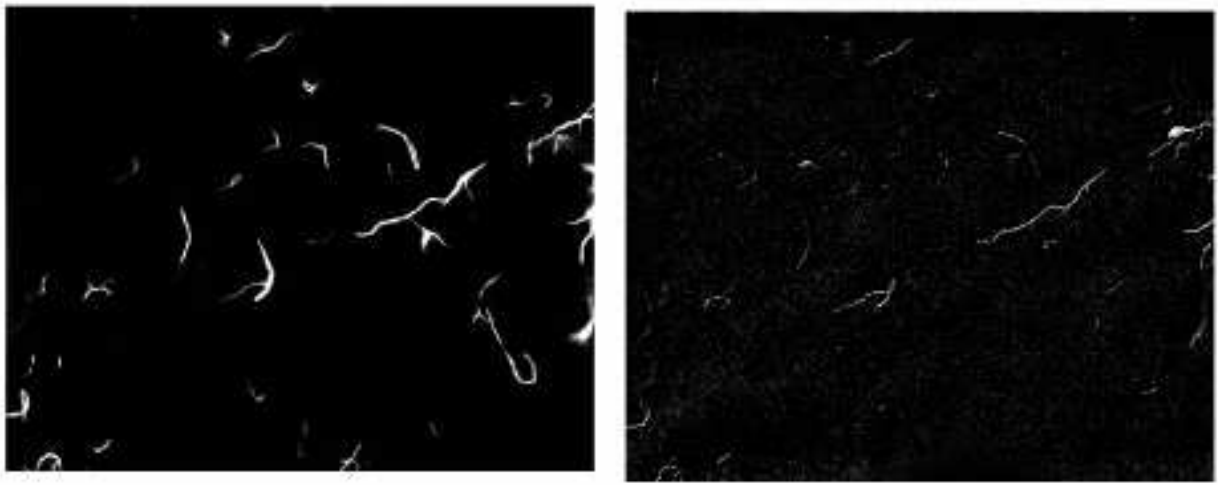}
\caption{ 
 \label{fig:SingleFA}
Left panel: 40X fluorescence image of  PEG~35000 bundles mixed with sparsely 
labeled (TRITC-phalloidin) F-actin.
Right panel: 40X DIC image at same location.
The sample thickness in these images is less than 10 $\mu$m, 
so bundles are much more sparse in these images than in Fig~\ref{fig:Nucl}.
Image size is 
$\rm{170~\mu m \times 135~\mu m}$.
}
\end{figure*}
 \hfill
 \newline

 It appears that individual labeled actin filaments show no size preference
when they combine with previously formed, unlabeled bundles.
When new phalloidin-labeled F-actin is introduced, 
the fluorescence seems to be uniformly distributed.
As is in Fig ~\ref{fig:SingleFA}, the DIC images co-localize with 
that of the TRITC fluorescence, suggesting that individual F-actin filaments
attach themselves to the previously formed bundles.
Also, it was possible to show that preformed bundles do
not exchange material,  as phalloidin-labeled bundles
mixed with unlabeled bundles show separate DIC and fluorescence
populations (data not shown).

\subsection{PEG is not part of the bundle structure}
As described in Methods, an elaborate procedure was designed to
address the question whether PEG is an integral part of the F-actin
bundle structure. 
Following a systematic set of measurements, it was concluded that less than 0.1\% w/w PEG was found to
be tightly bound or trapped with the bundles.
This result directly confirms that PEG-induced bundle
formation is the result of the depletion effect, rather than
a putative PEG binding or cross-linking of F-actin.

\section{Discussion}

 We interpret the data in Fig ~\ref{fig:C_o} as an interplay between the attractive
depletion force and the electrostatic repulsion of
two negatively charged cylinders in a monovalent salt solution.
We present approximations showing that it is reasonable
to expect the depletion force to cause F-actin to laterally aggregate.
Inclusion of a van der Waals attraction offers an explanation of the large hysteresis at 
high ionic strength,
although it is not clear exactly how much this interaction contributes to the final
bundled state. 

 In the following three sections we calculate the electrostatic, depletion, and
van der Waals potentials between two infinite, parallel F-actin cylinders in order to
understand the onset of bundle formation.
Based on the results of these calculations, issues such as the observed hysteresis 
and bundle size are discussed.

\subsection {Depletion Attraction}

  As described in the Introduction, the depletion layer and osmotic pressure create an 
additional free energy for a single colloidal particle in a polymer solution. If the layer thickness
were constant for all polymer concentrations and the osmotic pressure difference linear in concentration, 
this positive free energy 
would be $\int P \rm{d}V$,
where $P$ continuously passes
from $P_b$ at the colloid particle surface to zero in the surrounding solution.
Since this is not the case, 
the energy is found by integration over volume and concentration~\cite{tuinier_2001}
\begin{equation}
 W = \int_V \int_{C_b'=0}^{C_b'=C_b} \
 \rm{d}V {(1-{{C(\bf{r})} \over {C'_b}})}{{\partial P_b} \over {\partial C'_b}} \rm{d} C'_b   \label{PV_eqn}
\end{equation}
where $C(\bf{r})$ is the local polymer concentration; the bulk polymer
concentration is $C'_b$, and the local osmotic pressure 
$P_b$ is a function of $C'_b$.
This expression for integrating total free 
energy is in analogy with 
an electrostatic charging process to calculate, for
example, the energy required to charge a capacitor.
 
Following the qualitative reasoning of
the AO theory, the free energy for two interacting colloidal particles, as expressed in Eqn.~\ref{PV_eqn},
should become less positive as the two
approach, indicating an attractive depletion force. 
To calculate the interaction, one
places the two particles at separation $D$ and
zero polymer concentration. Then the polymer bulk concentration is
parametrically ramped to the final value $C_b$, with Eqn.~\ref{PV_eqn} used to evaluate $W$.
This process
has the qualitative and intuitively appealing
feature of generating a depletion force with a range of the order of
$R_g$ for all concentrations. 
We denote this change in free energy $W$ as the
particles approach from infinity as $U_D$.

 We employ osmotic pressure data and $\xi$ values from the PRISM theory~\cite{UIUC_2000}, and 
an ansatz about overlapping depletion regions~\cite{tuinier_2001} to estimate the
attractive depletion potential $U_D$.
For osmotic pressure $P(C)$, we adopt a virial equation $ P = B_1C + B_2 C^2$ from the
literature~\cite{UIUC_2000}.
The  values for $B_1$ and $B_2$ thus determined are presented in Table~\ref{tab:table1}.

 We use the proposal~\cite{tuinier_2001} that the polymer concentration in 
the overlap region $C(\bf{r}\it)$ can be expressed as a product of
the separate concentrations:
\begin{equation}
 C({\bf{r}}) = {{C_1({\bf{r}})} \over {C_b}} ~ {{C_2({\bf{r}})} \over {C_b}} C_b  \label{Creqn}
\end{equation}
where $C_b$ is the bulk concentration and $C_1$ and $C_2$ are the profiles surrounding the two colloids 
at infinite separation. 
$C({\bf{r}})$  is in effect the product of two probability densities. 
The concentration profile near an isolated 
cylinder is assigned a cone-shaped function of width 2 $\xi$:
\begin{equation}
C_i(r) = C_b {{r - R_A} \over {2 \xi}};~ R_A < r < R_A + 2 \xi \nonumber
\end{equation}

 With the virial expressions for $P(C_b)$ and the depletion profile $C(r)$ from Eqn.~\ref{Creqn},
we numerically evaluate Eqn.~\ref{PV_eqn}. 
Letting $C_b$ take
the values of $C_o$ in Fig ~\ref{fig:C_o} for PEG~35000, namely 1.6, 1.08, 0.97, and 0.80 \% w/w
(corresponding to 100~mM,  140~mM, 175~mM, and 500~mM KCl respectively),
we show the numerical results of  $U_D$ in Fig~\ref{fig:Ucalc}.
This more elaborate 
methods of calculating the attractive potential $U_D$ 
yield values  
less than the AO hard sphere model, typically
by a factor of five.

\begin{table}
\caption{\label{tab:table1}
A list of 4 parameters for PEG used in this study.
$B_1$ and $B_2$ are the first and second viral coefficients in
the formula
$ P = B_1C + B_2 C^2 $
where $C$ is percent w/w and the osmotic pressure
$P$ is in the unit of Pa.
The values for $R_g$ and $C^*$ are derived from Kulkanari \it{et. al.}\rm~\cite{UIUC_2000} 
and are consistent with other measurements~\cite{selser91}.  
}
\begin{ruledtabular}
\begin{tabular}{ccccc}
PEG MW & $B_1$ & $B_2$ & $R_g \rm{(nm)} $ & $C^* \rm{(\%w/w)} $ \\
\hline
8000 & 3112.5 & 771.1 & 4.7 & 3.0 \\
20000& 1245.0 & 641   & 8.2 & 1.4 \\
35000 & 711 & 574 & 11.4 & 0.9 \\

\end{tabular}
\end{ruledtabular}
\end{table}

\subsection {van der Waals attraction}

 The standard formula for the van der Waals interaction~\cite{israelachvili} for
two parallel cylinders, based on the Derjaguin approximation,
seemed to have an unreasonably long range, so a numerical integration of
the potential was done, based on the $1/{r^7}$ force between two
elements of identical material. 
Assuming two elements of material separated by $r$ experience
a force $ {F} =A \frac{6}{\pi^2} d^3 r_1 d^3 r_2 /{r_{12}^7} $ 
where $A$ is the Hamaker constant~\cite{overbeek_book}, 
we see that an element $dx_1 {d}y_1 {d}z_1 $ and a rod of
area $dx_2 dy_2 $
separated by $S$
experience a force $ F =A \frac{6}{\pi^2} \frac{5 \pi} {16} {d} x_1 {d} y_1 {d} z_1 {d}x_2 {d}y_2/{S^6}   $.
Considering a circular slice of F-actin, height  $dz$, and a parallel cylinder of F-actin
with lattice size $dz$, one can numerically calculate the total ${F} $
for a series of separations $D$, then numerically integrate $F {d}x$
to obtain a potential $ U_{\rm{vdW}}(D)$ for this slice. This function
is shown in Fig ~\ref{fig:Ucalc} for $A = 0.1~kT$, $L =100~\rm{nm}$.

\subsection{Electrostatic Repulsion}

\begin{figure}
\includegraphics{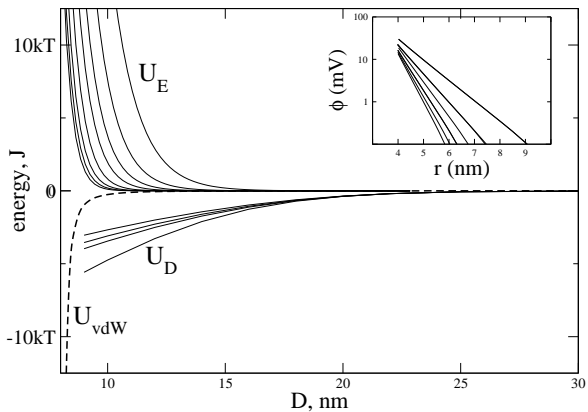}
\caption{
\label{fig:Ucalc}
Numerical values for $U_{\rm{vdW}}$,  $U_{D}$, and $U_{E}$.
For $U_{E}$, monovalent salt ranges from 100~mM to 600~mM.
For the $U_{D}$ family,  PEG~35000 concentration has the values
in  Fig ~\ref{fig:C_o}: 1.60, 1.08, 0.95, and 0.80 percent w/w. 
The cylinder radius and surface charge 
are those of fully charged F-actin, $R$ = 4~nm, $\lambda$ = ~4~e/nm,
For $U_{\rm{vdW}}$, a Hamaker constant $A = 0.1~kT$ was chosen.
$L = 100 \rm{nm}$.
Inset:  $\phi(r)$ for monovalent salt from 100~mM to 600~mM.
}

\end{figure}

 For a single cylinder of surface charge density $\sigma$, in 
salt solution of ionic strength $I$, the potential solution of  the
linearized Poisson-Boltzmann equation is known to be~\cite{zimm1,harries_98}
\begin{equation}
 \phi(r) = {\sigma K_0(\kappa r)} / {\epsilon \kappa K_1(R_A \kappa)} \label{PB1}
\end{equation}
where $K_0$ and  $K_1$ are modified Bessel functions,
and $\kappa^{-1}$ is the Debye screening length~\cite{overbeek_book};
for monovalent salts, $\rm{\kappa ^{-1} = 0.3~nm} / \sqrt I$ 
where $I$ is the value of ionic stength in moles/L.
The
numerical results described in Methods match closely Eqn.~\ref{PB1} for $I$ = 100~mM.
The inset of Fig ~\ref{fig:Ucalc} shows  numerical solutions $\phi(r)$ for KCl from 100 to 600~mM.

 We should mention in passing that with 
divalent counterions, and for cylinders of smaller diameter and higher surface charge,
$\phi(r)$ does not have the approximate form $ \phi = \phi_0 \exp(-(r-R_A)\kappa) $.
Under these conditions 
there is a sudden rise in potential as $r \rightarrow  R_A$.
For a cylinder of modest surface charge, the system free energy
is  $\phi_0 Q/2$~\cite{overbeek_book, sharp_90}.
In this situation, the surface potential is less than $kT/e=25 mV$, and
the positive entropic energy of the counterion cloud
is equal to its negative electrostatic energy~\cite{oosawa_polyele},
so change in system free energy is reflected as change in
surface potential. 
For F-actin in 100~mM KCl, the electrostatic potential at the surface
is calculated to be 30 mV (see inset of Fig~\ref{fig:Ucalc}). Therefore, F-actin with monovalent counter ions is in
the linear regime of the PB equation. In contrast, DNA (5.9 $e$/nm, 0.9~nm diameter)
has a  significantly higher surface potential, and  non-linear 
effects must be taken into account.

 We do not have a calculation of electrostatic interaction $U_E(D)$ for
two parallel cylinders in proximity. 
However, because the highest value for $\phi(r)$ (the surface potential) is not much greater 
than $kT/e$,
we use the sum $ \phi_1(r) + \phi_2(r)$ as an approximate
solution:
\begin{equation}
\nabla^2 (\phi_1 + \phi_2) \sim \sinh ( {{\phi_1 e} \over {kT}} + {{\phi_2 e} \over {kT}}) 
 \sim \sinh ( {{\phi_1 e} \over {kT}} )  + \sinh({{\phi_2 e} \over {kT}}) \nonumber
\end{equation}
We take the change in free energy as the two cylinders are
forced together to be like two charges in free space:
\begin{equation}
    U_E(D) = \int_{S_1} \phi_2 \sigma_1 dS_1    \label{U_E}
\end{equation}
For $D > 2R+\kappa^{-1}$ this is a very reasonable approximation~\cite{harries_98}.
We neglect the effect in the region where the dielectric of one cylinder
displaces the counterions of the other, assuming the 
protein cylinder has the same dielectric constant as the
surrounding medium.
In Fig ~\ref{fig:Ucalc} we show the numerical integration of
Eqn.~\ref{U_E} for [KCl] = 100 mM to 600 mM for fully charged F-actin.
These results match the low surface charge example of Harries~\cite{harries_98}.
 \hfill
 \newline
 \hfill
 \newline
 \hfill
 \newline
 \hfill
 \newline
 \hfill
 \newline

\subsection{Prediction of bundle formation}

  Our interpretation is that
the depletion potential is sufficient to allow the filaments to find a
lower energy condensed phase, which is normally denied by
Coulomb repulsion, a view consistent with
the standard DLVO theory. 
The qualitative feature to be appreciated is the role of ionic strength
in making the vdW interaction accessible. 

 \hfill
 \newline
 \hfill
 \newline
 \hfill
 \newline
 \begin{figure}
 \includegraphics{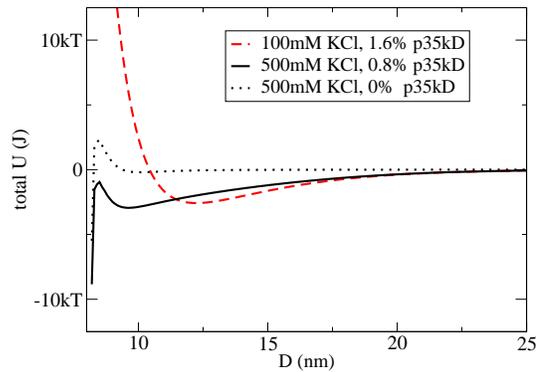} 
 \caption{
 \label{fig:U_tot}
Sum of $U_E$, $U_D$, and $U_{\rm{vdW}}$ for fully charged (4$e$/nm) F-actin.
Values for [KCl] and PEG-35kD are taken from the data of  Fig~\ref{fig:C_o}.
L = 100~nm.
}
 \end{figure}
 \hfill
 \newline

 Shown in  Fig ~\ref{fig:U_tot}  is the sum $U_D + U_E +U_{\rm{vdW}}$ for three  
representative conditions: 500~mM KCl, 0.8\% PEG~35kD; 100~mM KCl, 1.6\% PEG~35kD;
and 500~mM KCl, 0\% PEG. 
The calculations appear successful in explaining 
features of the results at 500~mM ionic strength, predicting bundle formation
with  0.8\% PEG~35kD, consistent with Fig~\ref{fig:C_o}. 
With  a rather low Hamaker constant, $A = 0.1~kT$, the electrostatic repulsion is 
large enough to stabilize the
suspension at 0\% PEG,  as it must. 
According to Eqn.~\ref{U_E}, 
fully charged F-actin in 100~mM KCl has
a large value of $U_E$ that totally prohibits the
role of van der Waals forces in the bound state.
This is contradictory with the experimental result.
Both 100~mM and 500~mM PEG 35kD bundles are similar in appearance, although hysteresis 
has a wider range for 500~mM KCl, as seen in Fig~\ref{fig:dissolve}. While the choice of 
the low Hamaker constant $A = 0.1~kT$ is an obvious cause of the imbalance, 
it is hard to justify using different values for Hamaker constant at different ionic strengths.
Additional mechanisms unaccounted for in our treatment are discussed below.

 Counterion fluctuation and redistribution~\cite{oosawa_polyele,liu_97,pincus_lau_02,ray_gsm_94} 
clearly play important roles in 
leading to an overall attractive interaction. These effects have not been taken into consideration in our 
simple electrostatic calculations. Since the effects are more pronounced at lower ionic strength, it 
is not surprising that our calculated $U_E$ based on PB theory overestimates the repulsion (Fig ~\ref{fig:Ucalc}).
at low ionic strength.
Most of these  works seem to  focus on polyelectrolytes of high surface charge density, such as
DNA, in the presence of  multivalent counterions. These theories may also be expanded
to treat systems of relatively weak surface charge, such as F-actin.

 The semi-quantitative fits to the data shown in Fig ~\ref{fig:C_o} suggest ligand binding 
as an alternative mechanism for predicting bundle formation.
With a mass action ligand (counterion) binding 
interpretation (Eqn.~\ref{K_A_eqn}) with $K_A \rm{ = 1/(5~mM)}$, 
half the negative charge
would be neutralized at 5~mM KCl, which implies a surface charge down by
a factor of 100 at 500~mM. 
This surface charge could certainly not support a stable suspension.
A weaker association constant, say $K_A$ = 1/(100~mM), fails to produce the flat baseline
of $C_o$ at higher salt. 
Experiments~\cite{carlier_86,selden_89}  focused mainly on F-actin's tightly bound divalent 
ions $\rm{Ca^{2+}}$ and
$\rm{Mg^{2+}}$ have demonstrated effects of $\rm{[K^+]}$ in the 
rapid phase of fluorescence induction, indicating low affinity $\rm{K^+}$ binding constants
from 1/(10~mM) to  1/(100~mM).
The data in Fig ~\ref{fig:C_o}, together with the fact that  F-actin 
is stable at the  highest monovalent ionic strength,  seem to demand that F-actin 
maintain a significant fraction of
unneutralized charge at high salt.
Indeed, the electrophoretic mobility measurements for filamentous phages fd and M13 suggest
 that the 
filaments remain charged up to 500 mM KCl [Q. Wen and J.X. Tang, unpublished data]. 

 A recent study by Yu and Carlsson~\cite{carlsson_03} directly addresses the
question of F-actin electrostatic interaction.
This work includes electrostatic and entropic terms 
in ligand (counterion) binding 
as well as the actual spacial configuration of the charged 
groups of the protein structure.
These authors point out a role of induced charge condensation as
the two filaments begin to interact: as the surface potential of  
one filament is increased by the other, the effective $K_A$ of each site is raised by
a factor $\exp(e \phi_2/kT)$, lowering the
repulsion by increasing the counterion binding. 
The actual interaction is found with a largest-error-correction
algorithm seeking the lowest energy, which is a function of
the charge of all sites.
Their treatment may be expanded to predict features
of PEG-induced bundle formation, which goes beyond
the scope of this work.

\subsection {Distinction from polyvalent cation induced bundle formation}
 The mechanism of  polymer-induced bundling is distinct from the 
bundling induced by divalent metal ions, cationic complexes, and
basic polypeptides~\cite{JXT97,JXT96,JXT-BJ-02}.
The most salient feature is the inhibiting nature of monovalent
salts for the latter \it{vs.}\rm   the enhancing effect for the former.
Secondly, polyvalent counterions such as basic polypeptides and  cationic complexes function
 as some binding agent in F-actin bundles.
An opposite role is held by PEG, which does not bind F-actin,  as confirmed 
in this study with the sedimentation assay using radio-labeled PEG. 
In addition, we show in this work that under some conditions the PEG-induced
bundles are morphologically different in size.

Another distinct property from polyvalent cation induced bundle formation
is that no resolubilization was found with high polymer concentration
or high ionic strength (2.5~M~KCl), in that the structures were stable
under these conditions. In contrast, for most colloidal systems, 
precipitation occurs with the addition of multivalent ions, only to go back
into suspension 
with increased concentration of the multivalent counterions.
Long known for some classical colloids, this has recently been shown for
the bacteriophages fd and M13 viruses~\cite{JXT-BJ-02}.
This phenomenon has been predicted for F-actin by recent
theoretical calculations~\cite{carlsson_03}.
However, experimental test of such a prediction has
yet to be performed.

There is an apparent multitude of proteins which either bind to F-actin,
affect the polymerization of actin, or enhance the lateral aggregation of F-actin~\cite{JXT97}. 
In view of the distinction between the polyvalent counterion induced 
bundling and the one by depletion effect shown in this work, it is helpful
to assess their respective contributions to the related biochemical functions. Under the 
unifying tie of electrostatic effects, however,
this work implies a strong relevance of polyelectrolyte physics to the observation 
that so many biochemically unrelated proteins bind F-actin and induce
formation of actin bundles.  Is the binding capability
due largely to the cylindrical geometry of actin filaments,
which tends to diminish electrostatic stability~\cite{zimm1,oosawa_polyele}?

\subsection {Hysteresis }

  The hysteresis in the dissolution of PEG induced actin bundles (Fig 3)
can be qualitatively explained by the following scenario:
At zero polymer, filament/filament contact is prohibited by the
electrostatic barrier. As the height of this barrier is overcome with
the depletion potential $U_D$ from increased  polymer concentration, F-actin is allowed into the
vdW binding configuration. If the polymer concentration is then lowered,
the tightly bound state remains stranded inside the $U_E$ barrier.
This can be seen in Fig ~\ref{fig:U_tot} for the case of 500~mM~KCl.

\subsection {What limits the bundle size? }

 Suppose that the total surface charge of a bundle
increases with size, while the surface charge density remains constant.
Recall that Eqn.~\ref{PB1} shows that the self-energy goes as $\lambda^2/\sqrt{I}R \sim \sigma^2 R/\sqrt{I}$. 
If the  surface charge density is a constant value (that of an individual F-actin), this
electrostatic self-energy per unit length of a bundle would increase as 
the bundle radius $R$.
Assuming the depletion potential to be of depth $\xi$, $U_{D}$ per unit length would scale as
$\xi^{1/2}R^{3/2}$, which is the shaded area of Fig ~\ref{fig:rod-rod} for small $\xi/R$. 
>From these considerations, there is no obvious limit in size because
$U_D$ outraces $U_E$ with increasing diameter $2R$. 
This simple scaling argument also predicts that linear
polyelectrolytes of larger diameter are more prone to bundling when 
other parameters are comparable.

 The final bundle structure is 
not necessarily a hexagonal array of rods, but rather it could be the product 
of diffusion limited aggregation~\cite{stanley_book}.
Such structures are expected to show surface roughness greater than the 
standard statistical deviation proportional
to the square root of the accumulation.  From Fig ~\ref{fig:SingleFA}
we know that an individual filament finds a low energy
configuration on a previously formed bundle. However, the
situation is different for two matured bundles, where, due to
surface contour,  the relative
interaction (contact) area fails to scale with diameter.
Such a structure is easily broken up by the 
turbulence of light mixing.
Thus there is a point in the hierarchy of assembly 
when the macroscopic hydrodynamic forces exceed the 
surface-surface interaction.
This argument explains why the
bundles do not collapse into one large mass, but
assume polydisperse structures which show no tendency to 
combine with each other after a certain level of self-assembly. 

The arguments above based on the surface roughness do not 
fully explain how bundles might reach some equilibrium size limit.
The experimental observations are inconclusive whether an 
equilibrium size distribution exists or is practically attainable.
PEG~8kD bundles appear to stop growing at smaller sizes (data not shown) than
those 
induced by PEG~35kD (Fig ~\ref{fig:Nucl} and Fig ~\ref{fig:SingleFA}), which clearly  
grow into large structures,  
their extent perhaps only limited by the progressively slow kinetics. 
In the test tube experiments, the effect of diffusion may be surpassed by the 
sedimentation effect due to gravity for the large bundles observed. Additional 
experiments are necessary in order to assess which effects are dominant 
in determining the actin bundle size.   

  The hysteresis data also implies an inhomogeneity, or variation in
'fitness', among the actin bundles.
Some of the structures 'survive' with reduced 
polymer while others have dissolved back to dispersed actin filaments. 
The inhomogeneity may be related to defects in the structure of
the bundles, or more simply a wide variation of the bundle sizes. Additional experiments 
may also be designed to test these different proposals. 

\subsection {Concluding remarks}

 Our calculations
are to be taken as reasonable estimates of the forces involved in bundle formation. More
exact calculations, especially of the electrostatic interaction,
presumably including a role of
mass action counter-charge accumulation, 
should help explain many aspects of the  results quantitatively. 
Nevertheless, 
results from the simple experiments reported here 
provide an opportunity to expand the understanding
of some basic problems in lateral aggregation of protein filaments.

\section {Acknowledgments}
 We wish to thank Prof.~David~Daleke for his generosity.
This work was supported by NSF-DMR~9988389, NIH-HL~67286, and by Indiana University.
\bibliography{peg-paper-Jan2}
\end{document}